\documentclass[twocolumn,showpacs,preprintnumbers,amsmath,amssymb]{revtex4}
\usepackage{graphicx}
\usepackage{dcolumn}
\usepackage{bm}
\usepackage{color}

\begin{document}

\title{Core polarization for the electric quadrupole moment of 
neutron-rich Aluminum isotopes}

\author{Kenichi Yoshida}
\affiliation{RIKEN Nishina Center for Accelerator-Based Science, Wako, Saitama 351-0198, Japan
}%

\date{\today}

\begin{abstract}
The core polarization effect 
for the electric quadrupole moment of 
the neutron-rich $^{31}$Al, $^{33}$Al and $^{35}$Al isotopes 
in the vicinity of the island of inversion are investigated by means of the 
microscopic particle-vibration coupling model in which the 
Skyrme Hartee-Fock-Bogoliubov and quasiparticle-random-phase approximation 
are used to calculate the single-quasiparticle wave functions and 
the excitation modes. 
It is found that 
the polarization charge for the proton $1d_{5/2}$ hole state 
in $^{33}$Al is quite sensitive to coupling to the neutrons in 
the $pf$-shell associated with the pairing correlations, 
and that 
the polarization charge in $^{35}$Al becomes larger due to the stronger collectivity 
of the low-lying quadrupole vibrational mode in the neighboring $^{36}$Si nucleus. 

\end{abstract}

\maketitle

\section{Introduction}
The nuclear structure far from the $\beta$-stability line 
has been studied very actively 
with the development of the new-generation 
radioactive-isotope beam techniques 
together with of 
the microscopic nuclear models 
applicable to drip-line nuclei 
carried out by the high performance computers. 

The ground state properties and the dynamical properties 
such as low-energy excitation modes and giant resonances in 
the medium-mass to the heavier nuclei 
have been successfully described by 
the self-consistent mean-field theory or 
the nuclear density-functional theory (DFT)~\cite{ben03}. 
The nuclear DFT has been applied to the exotic modes of excitation 
in unstable nuclei~\cite{vre05,paa07}, 
and developed towards description of nuclei in the whole chart. 
Along this line, 
the self-consistent random-phase approximation (RPA) 
including the pairing correlation and the nuclear deformation 
has been recently developed by several groups~\cite{pen07,per08,yos08}.

Presently, the small excitation energy of the $2_{1}^{+}$ state and 
large transition probability $B(E2;0^{+}\to 2_{1}^{+})$ in $^{32}$Mg 
has been discussed in connection with 
the breaking of the spherical magic number $N=20$ 
in neutron-rich systems~\cite{det79,gui84,mot95}. 
The recent gyromagnetic-factor measurement of $^{33}$Al at GANIL~\cite{him06} 
and the $\beta$-decay study of $^{33}$Mg at NSCL~\cite{tri08}
indicate that $^{33}$Al has a certain amount of $2p-2h$ intruder configuration. 

The electric quadrupole moment ($Q$ moment) 
representing the deviation from a sphere is directly related to 
the deformation property of the nucleus, 
and thus its investigation for neutron-rich nuclei at around $N=20$ 
is strongly desired both experimentally and theoretically~\cite{nag08}. 
Quite recently, the $Q$-moment measurement of $^{33}$Al has been performed 
at GANIL~\cite{nag09}.

In order to investigate 
the ground-state $Q$ moments of neutron-rich Al isotopes 
in the vicinity of the ``island of inversion"~\cite{war90}, 
we carry out the particle-vibration coupling (PVC) calculation 
based on the Skyrme density functional, 
on top of the self-consistent quasiparticle-RPA (QRPA). 

The article is organized as follows:
In Sec.~\ref{method}, the method is explained.
In Sec.~\ref{results}, 
we perform the numerical calculations and investigate 
the core polarization for the electric quadrupole moments in $^{31,33,35}$Al. 
Sec.~\ref{summary} contains the conclusions.

\section{\label{method}Method}
\subsection{Microscopic particle-vibration coupling model}
The nuclear Hamiltonian of a PVC model~\cite{BM2} 
on top of the Skyrme-Hartree-Fock-Bogoliubov (HFB) 
and QRPA is written as 
\begin{equation}
\hat{H}=\sum_{i}E_{i}\hat{\beta}^{\dagger}_{i}\hat{\beta}_{i}+
\sum_{\lambda}\hbar \omega_{\lambda} \hat{B}^{\dagger}_{\lambda} \hat{B}_{\lambda}
+\hat{H}_{\mathrm{couple}}.
\label{hami}
\end{equation}
Here $E_{i}$ is the quasiparticle energy obtained as a self-consistent solution 
of the Skyrme-HFB equation, 
$\hat{\beta}^{\dagger}_{i}, \hat{\beta}_{i}$ the quasiparticle creation and annihilation operators. 
The nucleon creation operator $\hat{\psi}^{\dagger}(\boldsymbol{r})$ is then represented 
using the quasiparticle wave functions as 
\begin{equation}
\hat{\psi}^{\dagger}(\boldsymbol{r})=\sum_{i}\varphi_{1,i}(\boldsymbol{r})\hat{\beta}^{\dagger}_{i}
+\varphi_{2,i}^{*}(\boldsymbol{r})\hat{\beta}_{i}.
\end{equation}
The phonon energy $\hbar \omega_{\lambda}$ is a solution of the QRPA equation 
on top of the Skyrme-HFB, and 
$\hat{B}^{\dagger}_{\lambda}, \hat{B}_{\lambda}$ 
the phonon creation and annihilation operators. 
We solve the Skyrme-HFB+QRPA equations in the $m$-scheme. 
Details of the calculation scheme is given in Ref.~\cite{yos08}.

Let us now consider the change of the density $\varrho(\boldsymbol{r})$ 
due to the collective vibrations as $\varrho(\boldsymbol{r})$ $\to$
$\varrho(\boldsymbol{r})+\delta \varrho(\boldsymbol{r},t)$. 
The nuclear potential $U[\varrho(\boldsymbol{r})]$ is accordingly changed as
$U[\varrho(\boldsymbol{r})]$ $\to$ $U[\varrho(\boldsymbol{r})+\delta \varrho(\boldsymbol{r},t)]$. 
To first order in the change of the density, 
the difference of the nuclear potential is evaluated to be
\begin{equation}
U[\varrho(\boldsymbol{r})+\delta \varrho(\boldsymbol{r},t)]-U[\varrho(\boldsymbol{r})]=
\int d\boldsymbol{r}^{\prime}
\dfrac{\delta U[\varrho(\boldsymbol{r})]}{\delta \varrho(\boldsymbol{r}^{\prime})}
\delta \varrho(\boldsymbol{r}^{\prime},t).
\end{equation}
Then, the PVC Hamiltonian has a form of
\begin{equation}
\hat{H}_{\mathrm{couple}}=\int d\boldsymbol{r} d\boldsymbol{r}^{\prime}
\dfrac{\delta U[\varrho(\boldsymbol{r})]}{\delta \varrho(\boldsymbol{r}^{\prime})}
\delta \varrho(\boldsymbol{r}^{\prime},t) 
\hat{\psi}^{\dagger}(\boldsymbol{r})
\hat{\psi}(\boldsymbol{r}).
\label{hami_pv}
\end{equation}

We introduce the vacuum defined by the product of the HFB vacuum and the QRPA vacuum;
\begin{equation}
\hat{\beta}_{i}|0\rangle=0, \hspace{1cm} \hat{B}_{\lambda}|0\rangle=0.
\end{equation}
The density variation $\delta \varrho(\boldsymbol{r},t)$ 
can be written in a second quantized form using the QRPA modes as 
\begin{equation}
\delta \hat{\varrho}(\boldsymbol{r})
=\sum_{\lambda}\left[ \delta \varrho_{\lambda}(\boldsymbol{r})\hat{B}^{\dagger}_{\lambda}
+\delta \varrho^{*}_{\lambda}(\boldsymbol{r})\hat{B}_{\lambda}\right],
\end{equation}
where $\delta \varrho_{\lambda}(\boldsymbol{r})$ 
is a transition density to the QRPA state $|\lambda\rangle=\hat{B}^{\dagger}_{\lambda}|0\rangle$.  

The $\hat{\beta}^{\dagger}_{i}\hat{\beta}^{\dagger}_{j}$ and 
$\hat{\beta}_{j}\hat{\beta}_{i}$ parts 
in $\psi^{\dagger}(\boldsymbol{r}^{\prime})\psi(\boldsymbol{r})$ 
are taken into account in the QRPA phonons~\cite{bri05}.  
Consequently, the PVC Hamiltonian in the leading order reads 
\begin{align}
\hat{H}_{\mathrm{couple}}=&\sum_{\lambda,ij}\int d\boldsymbol{r}d\boldsymbol{r}^{\prime}
\dfrac{\delta U[\varrho(\boldsymbol{r})]}{\delta \varrho(\boldsymbol{r}^{\prime})}
[\delta \varrho_{\lambda}(\boldsymbol{r}^{\prime})\hat{B}^{\dagger}_{\lambda}
+\delta \varrho^{*}_{\lambda}(\boldsymbol{r}^{\prime})\hat{B}_{\lambda}] \notag \\
&
\times [\varphi_{1,i}(\boldsymbol{r})\varphi^{*}_{1,j}(\boldsymbol{r})
-\varphi_{2,i}(\boldsymbol{r})\varphi^{*}_{2,j}(\boldsymbol{r})]\beta^{\dagger}_{i}\beta_{j}.
\label{hami_pv3}
\end{align}
The higher order effects can be treated systematically in the Nuclear Field Theory~\cite{bor77}.

The coupling interaction in Eq.~(\ref{hami_pv3}) is derived from the Skyrme density functional. 
In the present calculation, we approximate the momentum dependent 
terms in the Skyrme interaction by the Landau-Migdal (LM) form. 
This approximation is made only for the construction of 
the PVC Hamiltonian as in Refs.~\cite{sag84,ham96}. 
The isoscalar (IS) and the isovector (IV) coupling interactions are expressed as
\begin{equation}
\dfrac{\delta U[\varrho(\boldsymbol{r})]}{\delta \varrho(\boldsymbol{r}^{\prime})}
\delta \varrho_{\lambda}(\boldsymbol{r}^{\prime})
=
\begin{cases}
v^{\tau=0}(\boldsymbol{r})\delta \varrho_{\lambda}^{\mathrm{IS}}(\boldsymbol{r}^{\prime})
\delta(\boldsymbol{r}-\boldsymbol{r}^{\prime}) \\
v^{\tau=1}(\boldsymbol{r})\delta \varrho_{\lambda}^{\mathrm{IV}}(\boldsymbol{r}^{\prime})
\delta(\boldsymbol{r}-\boldsymbol{r}^{\prime})\tau_{z}\tau^{\prime}_{z}.
\end{cases}
\end{equation}
The explicit expressions for $v^{\tau=0}(\boldsymbol{r})=F_{0}/N_{0}$ 
and $v^{\tau=1}(\boldsymbol{r})=F_{0}^{\prime}/N_{0}$ 
are given in Ref.~\cite{gia81}.

\subsection{Description of odd-$A$ systems}
In order to describe the odd-$A$ nuclear systems, 
we diagonalize the Hamiltonian (\ref{hami}) 
within the subspace 
$\{\hat{\beta}^{\dagger}_{i}|0\rangle, \hat{B}^{\dagger}_{\lambda}\hat{\beta}^{\dagger}_{j}|0\rangle \}$. 
Then, the resulting state vector is written as
\begin{equation}
|\phi\rangle=\sum_{i}c_{i}^{0}\hat{\beta}^{\dagger}_{i}|0\rangle
+\sum_{\lambda j}c^{1}_{\lambda j}\hat{B}^{\dagger}_{\lambda}\hat{\beta}^{\dagger}_{j}|0\rangle.
\end{equation}

The operator for the quadrupole moment can be written as
\begin{equation}
\hat{Q}=\langle \hat{Q}\rangle +
\sum_{ij\in \pi}Q_{ij}\hat{\beta}^{\dagger}_{i}\hat{\beta}_{j}
+\sum_{\lambda}(Q_{\lambda}\hat{B}^{\dagger}_{\lambda}+Q_{\lambda}^{*}\hat{B}_{\lambda}),
\end{equation}
where
\begin{subequations}
\begin{align}
Q_{ij}&=\langle 0|\hat{\beta}_{i}\hat{Q}\hat{\beta}^{\dagger}_{j}|0\rangle \notag \\
&=\int d\boldsymbol{r} (3z^{2}-r^{2})
[\varphi_{1,i}(\boldsymbol{r})\varphi^{*}_{1,j}(\boldsymbol{r})-
\varphi_{2,i}(\boldsymbol{r})\varphi^{*}_{2,j}(\boldsymbol{r})], \\
Q_{\lambda}&=\langle 0|[\hat{B}_{\lambda},\hat{Q}]|0\rangle 
=\int d\boldsymbol{r} (3z^{2}-r^{2})\delta\varrho_{\lambda}^{\pi}(\boldsymbol{r}),
\end{align}
\end{subequations}
and $\langle \hat{Q}\rangle$ is the vacuum expectation value.

The electric $Q$ moment of the eigenstate $|\phi \rangle$ is then calculated as
\begin{align}
\langle \phi|e\hat{Q}|\phi\rangle
=&e \Bigl\{ \langle \hat{Q}\rangle+\sum_{i}\left[(c^{0}_{i})^{2}Q_{ii}
+2c^{0}_{i}\sum_{\lambda}c^{1}_{\lambda i}Q_{\lambda}\right] \notag \\
&+\sum_{\lambda,jk}c^{1}_{\lambda j}c^{1}_{\lambda k}Q_{jk} \Bigr\}.
\label{Q_mom}
\end{align} 

We apply this model to odd-$Z$ nuclei to calculate  
the proton polarization charge of the state $|i\rangle$ 
for the $Q$ moment, which is defined as
\begin{equation}
e^{\pi}_{\mathrm{pol}}=
e \left(\dfrac{\langle \phi|\hat{Q}|\phi\rangle}{\langle i|\hat{Q}|i \rangle}-1 \right),
\label{charge}
\end{equation}
where 
$\langle i |\hat{Q}| i\rangle=\langle \hat{Q}\rangle+Q_{ii}$.

\subsection{Parameters}
For the mean-field Hamiltonian, we employ the SkM* interaction~\cite{bar82}
in the present numerical applications.
We use the lattice mesh size $\Delta\rho=\Delta z=0.6$ fm 
and a box
boundary condition at ($\rho_{\mathrm{max}}=9.9$ fm, $z_{\mathrm{max}}=9.6$ fm).
The quasiparticle energy cutoff is chosen at $E_{\mathrm{qp,cut}}=60$ MeV
and the quasiparticle states up to $\Omega^{\pi}=15/2^{\pm}$ are included.
The pairing strength parameter is
determined so as to reproduce the experimental pairing gap for neutrons in 
$^{34}$Mg ($\Delta_{\mathrm{exp},\nu}=1.7$ MeV) obtained by the
three-point formula~\cite{sat98}. 
The strength $t_{0}^{\prime}=-295$ MeV fm$^{3}$ for the
mixed-type pairing interaction with 
the exponent of the density-dependence $\gamma=1$ leads to the pairing gap 
$\langle \Delta_{\nu}\rangle=1.71$ MeV in $^{34}$Mg~\cite{yos08}. 
On top of the Skyrme-HFB, 
we solve the QRPA equation within the space of the two-quasiparticle excitation 
of $E_{\alpha}+E_{\beta}\le 60$ MeV. 
The momentum-dependent terms in the residual interaction are exactly treated. 

The density-dependent Landau parameters $F_{0}$ and $F_{0}^{\prime}$ 
for the PVC interaction 
are determined by the parameters of the SkM* interaction. 
As shown in Refs.~\cite{yos08,miz07}, the attraction of the LM interaction is 
stronger than that of the self-consistent interaction in which 
the momentum-dependent terms are treated exactly. 
Therefore, we multiply the overall factor $f_{\mathrm{LM}}$ 
for the PVC interaction. 
We use $f_{\mathrm{LM}}=0.8$ and 0.9 for comparison. 
Accordingly, 
the difference of the results can be considered as a theoretical uncertainty.

\section{\label{results}Results and discussion}
\subsection{Properties of $^{32,34,36}$Si}
We describe the odd-$Z$ neutron-rich Al isotopes as 
a proton single-hole state coupled to the neighboring Si isotopes 
because the pairing gaps of protons in Si isotopes are zero. 
We summarize here the ground state properties and 
the structure of quadrupole excitations in $^{32,34,36}$Si. 

\begin{table}[b]
\begin{center}
\caption{Ground state properties of $^{32,34,36}$Si 
obtained by the deformed HFB calculation 
with the SkM* interaction and the mixed-type pairing interaction. 
Chemical potentials, 
average pairing gaps, root-mean-square radii for 
neutrons and protons are listed. 
The average pairing gaps of protons are zero in these isotopes. 
The average pairing gap is defined 
$\langle \Delta \rangle_{q}=-\int d\boldsymbol{r}\tilde{h}\tilde{\varrho}/\int d\boldsymbol{r}\tilde{\varrho}$.}
\label{GS}
\begin{tabular}{cccc}
\hline \hline\noalign{\smallskip}
 & $^{32}$Si & $^{34}$Si & $^{36}$Si   \\
\noalign{\smallskip}\hline\noalign{\smallskip}
$\lambda_{\nu}$ (MeV) & $-7.76$ & $-6.60$ & $-5.51$  \\
$\lambda_{\pi}$ (MeV) & $-13.5$ & $-15.9$ & $-17.3$  \\
$\langle \Delta \rangle_{\nu}$ (MeV) & 1.56 & 1.67 & 1.94 \\
$\sqrt{\langle r^{2} \rangle_{\nu}}$ (fm) & 3.22 & 3.32 & 3.39  \\
$\sqrt{\langle r^{2} \rangle_{\pi}}$ (fm) & 3.10 & 3.13 & 3.16  \\
\noalign{\smallskip}\hline \hline
\end{tabular}
\end{center}
\end{table}

In Table~\ref{GS}, the ground state properties are summarized. 
The neutron-rich Si isotopes under investigation are spherical 
although the calculated deformation parameters are not exactly zero 
($\beta_{2}=0.02$ in $^{34}$Si). 
This is due to the artificial breaking of the spherical symmetry 
associated with the finite mesh size and the rectangular box, 
and thus it is considered as a numerical error. 
The average pairing gaps of neutrons are finite, while 
those of protons are zero. 
This indicates that the $^{34}$Si 
has a neutron $2p-2h$ configuration in its ground state. 
The neutron occupation number of the $1f_{7/2}$ orbital is 
0.31, 0.78 and 2.21 in $^{32,34,36}$Si and 
that of the $2p_{3/2}$ orbital is 0.10 in $^{36}$Si.

\begin{figure}[t]
\begin{center}
\includegraphics[scale=1.0]{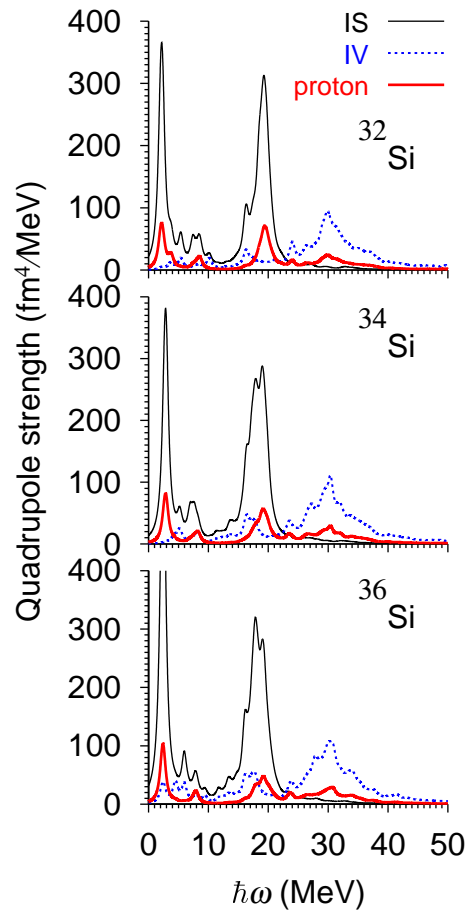}
\caption{(Color online) Response functions for the 
isoscalar (IS), isovector (IV) and proton quadrupole 
excitations in $^{32,34,36}$Si. 
The transition strengths are smeared by a Lorentzian
function with a width of $\Gamma=1$ MeV.}
\label{Si_response}
\end{center}
\end{figure}

Figure~\ref{Si_response} shows the response functions for 
the isoscalar (IS), isovector (IV) and proton quadrupole excitations. 
We can see a prominent peak at around 3 MeV in all of the isotopes under investigation. 
The isoscalar transition strengths are $B(\mathrm{IS}2;0^{+}\to 2_{1}^{+})=626, 637$ and 1137 fm$^{4}$ 
in $^{32,34,36}$Si, corresponding to 104, 98 and 162 in Weisskopf unit. 
Besides the low-lying collective $2^{+}$ state, 
we can see the giant quadrupole resonances (GQR) 
at around 20 MeV and 30 MeV for the IS and IV excitations. 

The $2_{1}^{+}$ state in $^{32}$Si 
is mainly generated by the $sd$-shell configurations of neutron and proton. 
The microscopic structure of the $K^{\pi}=0^{+}$ component is given by 
the two-quasiparticle excitations of $(\nu 2s_{1/2}\otimes 1d_{3/2})$ with a weight of 0.04,  
$(\nu 1d_{3/2})^{2}$ with 0.61, and  
$(\pi 2s_{1/2}\otimes 1d_{5/2})$ with 0.30. 
The microscopic structure of the $K^{\pi}={1}^{+}$ and $2^{+}$ components 
is the same within the numerical accuracy 
as of the $K^{\pi}=0^{+}$ component 
because of the spherical symmetry. 
The strength in the energy region $15\le \hbar\omega \le 25$ MeV exhausts 
79.4\% of the IS energy-weighted sum rule (EWSR) value. 
The IS strength in the low energy region up to 10 MeV exhausts 1.2\% of EWSR. 
The IV strength is distributed in a wider energy range $20 \le \hbar \omega \le 40$ MeV. 
The summed strength in this energy region exhausts 75.2\% of the IV-EWSR value.

In $^{34}$Si 
the neutron excitations into the $pf$-shell become appreciable 
for the $2_{1}^{+}$ state. 
The microscopic structure of the $K^{\pi}=0^{+}$ component is given by 
the $(\nu 1d_{3/2})^{2}$ excitation with a weight of 0.21, 
$(\nu 1f_{7/2})^{2}$ with 0.10, and 
$(\pi 2s_{1/2}\otimes 1d_{5/2})$ with 0.63.
The strength in the energy region $15\le \hbar\omega \le 25$ MeV exhausts 
79.8\% of the IS-EWSR value, 
and the strength in the energy region $20 \le \hbar \omega \le 40$ MeV 
exhausts 74.3\% of the IV-EWSR value. 

The $K^{\pi}=0^{+}$ component of the $2_{1}^{+}$ state in $^{36}$Si 
is mainly generated by 
the $(\nu 1f_{7/2})^{2}$ excitation with a weight of 0.39, 
$(\nu 2p_{3/2})^{2}$ with 0.08, and 
$(\pi 2s_{1/2}\otimes 1d_{5/2})$ with 0.47.
The strength in the energy region $15\le \hbar\omega \le 25$ MeV exhausts 
77.6\% of the IS-EWSR value, 
and the strength in the energy region $20 \le \hbar \omega \le 40$ MeV 
exhausts 72.9\% of the IV-EWSR value. 
Due to the mixing of the IS and IV modes in neutron-rich nuclei, 
we can see an appreciable IV strength in the lower energy region 
$10 \le \hbar \omega \le 20$ MeV. 
The summed strength in this energy region exhausts 10.4\% of the IV-EWSR value.

\subsection{Polarization charges in $^{31,33,35}$Al}
For describing the $I^{\pi}=5/2^{+}$ state of $^{31,33,35}$Al, 
we diagonalize the Hamiltonian (\ref{hami}) in the model space of 
the proton single-hole state of the $1d_{5/2}$ orbital $|(\Omega^{\pi}=-5/2^{+})^{-1}\rangle$ 
and the coupled states of 
$|(\Omega^{\pi}=-5/2^{+})^{-1}\otimes \omega_{K=0}\rangle$, 
$|(-3/2^{+})^{-1}\otimes \omega_{K=1}\rangle$ and $|(-1/2^{+})^{-1}\otimes \omega_{K=2}\rangle$.
We take the QRPA states $|\omega_{\lambda}\rangle$ 
whose IS or IV quadrupole transition strengths possessing greater than in 1 W.u. 

The dimension of the Hamiltonian (\ref{hami}) is 225 for $^{31}$Al. 
The $I^{\pi}=5/2^{+}$ state of $|^{31}\mathrm{Al} \rangle$ is 
constructed mainly by the one hole state and 
the hole coupled to the $2_{1}^{+}$ state as
\begin{align}
|{}^{31}\mathrm{Al}; I^{\pi}=5/2^{+},& M=5/2 \rangle = 0.93|(-5/2^{+})^{-1}\rangle \notag \\ 
& +0.21 |(-5/2^{+})^{-1}\otimes 2^{+}_{1}(K=0)\rangle \notag \\
& -0.23 |(-3/2^{+})^{-1}\otimes 2_{1}^{+}(K=1)\rangle \notag \\
& +0.16 |(-1/2^{+})^{-1}\otimes 2_{1}^{+}(K=2)\rangle. \label{31Al_wf}
\end{align}
The amplitude associated with the other components are smaller than 0.1. 
The coupled states in Eq.~(\ref{31Al_wf}) 
correspond to the $|(\pi 1d_{5/2})^{-1}\otimes 2_{1}^{+}\rangle$ state 
in the $j$-scheme representation. 
The ratios of the amplitudes are identical to those of 
the Clebsch-Gordan coefficients 
$\langle \cfrac{5}{2} \cfrac{5}{2} 2 0 |\cfrac{5}{2} \cfrac{5}{2}\rangle$, 
$\langle \cfrac{5}{2} \cfrac{3}{2} 2 1 |\cfrac{5}{2} \cfrac{5}{2}\rangle$ and 
$\langle \cfrac{5}{2} \cfrac{1}{2} 2 2 |\cfrac{5}{2} \cfrac{5}{2}\rangle$.
The quadrupole moment of $^{31}$Al is then 
calculated using Eq.~(\ref{Q_mom}) as 
$13.6 (13.9) e \mathrm{fm}^{2}$, 
where we use $f_{\mathrm{LM}}=0.8$ (0.9). 
From this value, the polarization charge of Eq.~(\ref{charge}) is calculated 
as $e^{\pi}_{\mathrm{pol}}=1.03e (1.08e)$. 
The result of the calculation overestimates slightly 
the experimental value of the $Q$ moment $11.2\pm 3.2 e\mathrm{fm}^{2}$~\cite{nag08}. 

For $^{33}$Al, the dimension of the Hamiltonian (\ref{hami}) is 250. 
As in $^{31}$Al, the wave function of $^{33}$Al is written 
by mainly the one hole state with a weight (the squared amplitude) of 0.90 
and the hole coupled to the $2_{1}^{+}$ state with a weight of 0.08.
The calculated values of the $Q$ moment and the polarization charge are 
$13.0 (13.4) e \mathrm{fm}^{2}$ and $e^{\pi}_{\mathrm{pol}}=0.89e (0.96e)$, respectively. 
The quadrupole moments and polarization charges of $^{31}$Al and $^{33}$Al 
are not very different. 

The dimension of the Hamiltonian (\ref{hami}) is 285 for $^{35}$Al. 
As in the case of $^{31}$Al and $^{33}$Al, 
the wave function of $^{35}$Al is written 
by the one hole state with a weight of 0.85 and 
the hole coupled to the $2_{1}^{+}$ state with 0.14. 
The contribution of the coupling to the $2_{1}^{+}$ state is larger 
than in $^{31,33}$Al. 
The calculated values of the $Q$ moment and the polarization charge are 
$14.7 (15.1) e \mathrm{fm}^{2}$ and $e^{\pi}_{\mathrm{pol}}=1.12e (1.18e)$, respectively. 

In order to see separately the effects of 
coupling to the low-lying modes and to the giant resonances 
on the polarization charge, 
we diagonalize the Hamiltonian (\ref{hami}) in the model space 
containing only the RPA modes with their energies larger than 10 MeV. 
The obtained polarization charges for the $1d_{5/2}$ orbital are 
$e^{\pi}_{\mathrm{pol}}=0.20e (0.22e)$, 
$0.21e (0.24e)$ and $0.18e (0.20e)$ in  $^{31}$Al, $^{33}$Al and $^{35}$Al, respectively. 
These values are close to the systematic value obtained in Ref.~\cite{suz03} 
($0.21e, 0.18e$ and $0.15e$ in $^{31,33,35}$Al), 
where the microscopic PVC calculations were performed including only the giant resonances 
on top of the self-consistent HF+RPA in light neutron-rich nuclei. 

The enhancement of the polarization charge in $^{35}$Al is thus 
due to the strong collectivity of the low-lying quadrupole vibrational mode 
in the core nucleus $^{36}$Si 
because the effects of coupling to the giant resonances are not 
sensitive to the neutron number in Al isotopes under investigation. 

\begin{figure}[t]
\begin{center}
\includegraphics[scale=0.65]{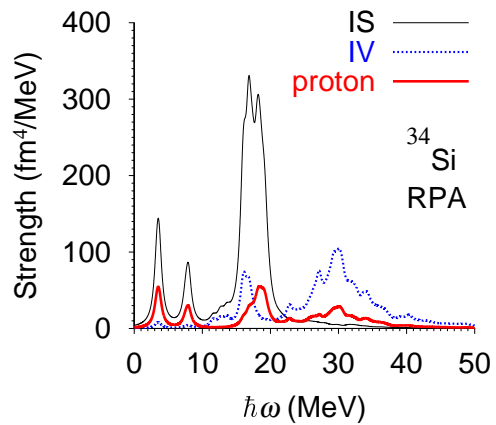}
\caption{(Color online) Same as Fig.~\ref{Si_response} 
but for the RPA strengths in $^{34}$Si.}
\label{34Si_response}
\end{center}
\end{figure}

\subsection{Effects of the pairing correlations in $^{33}$Al}

We investigate the effects of coupling to the $pf$-shell in $^{33}$Al. 
The dominant correlation in the present case is the pairing correlation 
because the core nucleus $^{34}$Si is calculated to be spherical at the HFB level. 

Figure~\ref{34Si_response} shows the IS, IV and proton quadrupole transition 
strengths in $^{34}$Si obtained by solving the Skyrme HF+RPA equations 
without pairing correlations. 
The collectivity of low-lying states are tremendously weakened, 
while the structure of GQR is not very different to that obtained by 
solving the Skyrme HFB+QRPA equations shown in Fig.~\ref{Si_response}. 

The $2_{1}^{+}$ state is constructed dominantly by 
the $(\pi 2s_{1/2}\otimes 1d_{5/2})$ excitation with a weight of 0.98. 
The strength $B(\mathrm{IS}2;0^{+}\to 2_{1}^{+})$ has only 237 fm$^{4}$. 
Using the RPA transition densities, we diagonalize the Hamiltonian (\ref{hami}) 
with a dimension of 241. 
The calculated wave function is mainly generated by 
the one hole state with a weight of 0.97 and 
the hole coupled to the $2_{1}^{+}$ state with a weight of 0.02.
This state is thus dominantly described by the proton $sd$-shell configurations 
because the $2_{1}^{+}$ state is generated by the proton excitation to 
the $2s_{1/2}$ orbital.

The resulting $Q$ moment is
$11.0 (11.4) e\mathrm{fm}^{2}$, 
and the polarization charge is $e^{\pi}_{\mathrm{pol}}=0.59e (0.65e)$. 

The $Q$ moment and the polarization charge in $^{33}$Al 
are quite sensitive to the neutron pairing correlation at $N=20$. 
This is similar to the enhancement mechanism 
of the $B(E2;0^{+}\to 2_{1}^{+})$ in $^{32}$Mg 
because the neutron pairing correlation is indispensable for 
the strong collectivity of the $2_{1}^{+}$ state in $^{32}$Mg~\cite{yam04}.

\subsection{$^{32}$Mg as a core}
\begin{figure}[t]
\begin{center}
\includegraphics[scale=0.65]{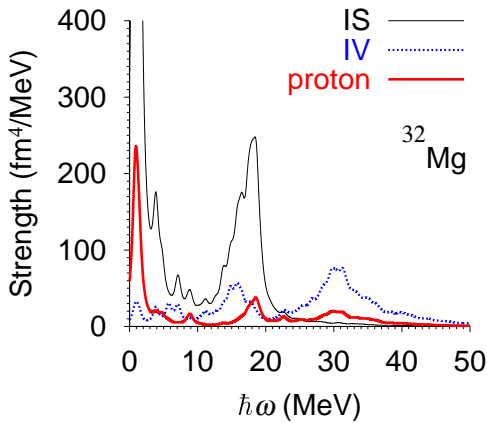}
\caption{(Color online) Same as Fig.~\ref{Si_response} 
but in $^{32}$Mg.}
\label{32Mg_response}
\end{center}
\end{figure}

The wave function $|^{33}$Al$\rangle$ can also be constructed 
by the quasi-proton coupled to $^{32}$Mg. 
The pairing gaps of neutrons and protons are 1.80 and 1.42 MeV. 
Figure~\ref{32Mg_response} shows the response functions for 
the IS, IV and proton quadrupole excitations. 
At 0.9 MeV, we can see a prominent peak possessing 
2414 fm$^{4}$ and 430 $e^{2}$fm$^{4}$ for the IS quadrupole strength 
and $B(E2)$, respectively. 
This result agrees well the experimental value~\cite{mot95}. 
The $2_{1}^{+}$ state is constructed by the two-quasiparticle excitations of 
$(\nu 1d_{3/2})^{2}$ with a weight of 0.10, 
$(\nu 1f_{7/2})^{2}$ with 0.09, 
$(\pi 1d_{5/2}\otimes 2s_{1/2})$ with 0.11, and 
$(\pi 1d_{5/2})^{2}$ with 0.61. 

We diagonalize the Hamiltonian (\ref{hami}) 
within the space of the proton quasiparticle of the $1d_{5/2}$ orbital, 
and the coupled states of a $sd$-shell quasi-proton to the quadrupole modes in $^{32}$Mg. 
In the present calculation with a dimension of 276, 
$|^{33}$Al$\rangle$ is constructed by 
the quasi-proton of the $1d_{5/2}$ level with a weight of 0.85, 
$|\pi 1d_{5/2}\otimes 2_{1}^{+}\rangle$ with 0.10, 
$|\pi 2s_{1/2}\otimes 2_{1}^{+}\rangle$ with 0.04, and 
$|\pi 1d_{3/2}\otimes 2_{1}^{+}\rangle$ with 0.01. 
The electric $Q$ moment of $^{33}$Al is then calculated as 12.4 (12.6) $e$fm$^{2}$. 
This is consistent with the calculation in Sec.~III.B. 

\section{\label{summary}Conclusion}
The polarization charges for the electric quadrupole moment of the 
neutron-rich Al isotopes at around $N=20$ have been investigated 
by carrying out the 
microscopic particle-vibration coupling calculation 
in which the coordinate-space Skyrme-Hartree-Fock-Bogoliubov and 
quasiparticle-random-phase approximation are employed 
to calculate the single-quasiparticle wave functions 
and the transition densities.  

It has been found that the neutron pairing correlations are crucial 
to generate collectivity of the $2_{1}^{+}$ state in $^{34}$Si, 
and that the polarization charge of the proton hole state of the $1d_{5/2}$ orbital 
becomes small in the absence of the 
pairing correlation in $^{33}$Al. 
The effect of 
the neutron pairing correlation at $N=20$ on 
the enhancement of the polarization charge in $^{33}$Al 
is very similar to the enhancement mechanism of the $B(E2)$ in $^{32}$Mg~\cite{yam04}. 

Effects of coupling to the giant resonances on the polarization charge 
are not very different in a small region of isotopes, 
and the low-lying collective modes have much 
effect on the polarization charge. 
Therefore,  
the polarization charge in $^{35}$Al is larger than in $^{31,33}$Al 
as a consequence of the stronger collectivity of the $2_{1}^{+}$ state 
in $^{36}$Si.  

\begin{acknowledgments}
The author acknowledges K.~Matsuyanagi, T.~Nakatsukasa for 
valuable discussions and encouragement, and 
T.~Nagatomo, H.~Ueno for stimulating discussions. 
He is supported by the Special Postdoctoral Researcher Program of RIKEN.
The numerical calculations were performed on the NEC SX-8 supercomputer
at the Yukawa Institute for Theoretical Physics, Kyoto University and 
the NEC SX-8R supercomputer 
at the Research Center for Nuclear Physics, Osaka University.
\end{acknowledgments}

\begin{appendix}
\end{appendix}


\begin{thebibliography}{99}
\bibitem{ben03}
M.~Bender and P.~-H.~Heenen, 
Rev. Mod. Phys. {\bf 75}, 121 (2003).

\bibitem{vre05}
D.~Vretenar, A.~V.~Afanasjev, G.~A.~Lalazissis and P.~Ring,
Phys. Rep. {\bf 409}, 101 (2005).

\bibitem{paa07}
N.~Paar, D.~Vretenar, E.~Khan and G.~Col\`o,
Rep. Prog. Phys. {\bf 70}, 691 (2007).

\bibitem{pen07}
D.~Pe\~na Arteaga and P.~Ring, 
Prog. Part. Nucl. Phys. {\bf 59}, 314 (2007).

\bibitem{per08}
S.~P\'eru and H.~Goutte, Phys. Rev. C {\bf 77}, 044313 (2008).

\bibitem{yos08}
K.~Yoshida and N.~Van Giai, Phys. Rev. C {\bf 78}, 064316 (2008).


\bibitem{det79}
C.~D\'etraz {\it et al}., 
Phys. Rev. C {\bf 19}, 164 (1979).
 
\bibitem{gui84}
D.~Guillemaud {\it et al}., 
Nucl. Phys. {\bf A426}, 37 (1984).

\bibitem{mot95}
T.~Motobayashi {\it et al}., 
Phys. Lett. {\bf B346}, 9 (1995).

\bibitem{him06}
P.~Himpe {\it et al}., 
Phys. Lett. {\bf B643}, 257 (2006).

\bibitem{tri08}
V.~Tripathi {\it et al}., 
Phys. Rev. Lett. {\bf 101}, 142504 (2008).

\bibitem{nag08}
D.~Nagae {\it et al.,} 
arXiv: 0819.2879. 

\bibitem{nag09}
T.~Nagatomo {\it et al}., 
{\it Proceedings of the 5th International Conference on Exotic Nuclei and Atomic Masses}, 
Ryn, Poland, 7 - 13 September, 2008, 
Eur. Phys. J. A (submitted). 

\bibitem{war90}
E.~K.~Warburton, J.~A.~Becker, and B.~A.~Brown, Phys. Rev. C {\bf 41}, 1147 (1990). 

\bibitem{BM2}
A.~Bohr and B.~R.~Motteleson, 
{\it Nuclear Structure}, vol.~II (Benjamin, 1975; World Scientific, 1998).

\bibitem{bri05}
D.~M.~Brink and R.~A.~Broglia, {\it Nuclear Superfluidity, 
Pairing in Finite Systems} 
(Cambridge University Press, 2005).

\bibitem{bor77}
P.~F.~Bortignon, R.~A.~Broglia, D.~R.~Bes and R.~Liotta, Phys. Rep. {\bf 30}, 305 (1977).

\bibitem{sag84}
H.~Sagawa and B.~A.~Brown, Nucl. Phys. {\bf A430}, 84 (1984). 

\bibitem{ham96}
I.~Hamamoto and H.~Sagawa, Phys. Rev. C {\bf 54}, 2369 (1996). 

\bibitem{gia81}
N.~Van Giai and H.~Sagawa, Phys. Lett. {\bf B106}, 379 (1981).

\bibitem{bar82}
J.~Bartel, P.~Quentin, M.~Brack, C.~Guet and H.-B.~H\r{a}kansson,
Nucl. Phys. {\bf A386}, 79 (1982).

\bibitem{sat98}
W.~Satu{\l}a, J.~Dobaczewski and W.~Nazarewicz, Phys. Rev. Lett. {\bf 81}, 3599 (1998).

\bibitem{miz07}
K.~Mizuyama, M.~Matsuo and Y.~Serizawa, arXiv: 0706.1115.

\bibitem{suz03}
T.~Suzuki, H.~Sagawa and K.~Hagino, Phys. Rev. C {\bf 68}, 014317 (2003).

\bibitem{yam04}
M.~Yamagami and N.~Van Giai, Phys. Rev. C {\bf 69}, 034301 (2004). 

\end{thebibliography}
\end{document}